\documentclass[final]{aipproc}

\layoutstyle{6x9}

\begin{document}

\title{High $p_t$ Suppression in CuCu Collsions at $\sqrt{s_{NN}} = 200$ GeV at RHIC}



\author{Selemon Bekele for the BRAHMS Collaboration}{
  address={University of Kansas, Lawrence, Ks 66045}
}

\begin{abstract}
Collisions between hadronic systems at relativistic energies provide
a window on the small-x gluon distributions of fast moving nuclei. It has been
predicted that gluon saturation effects will manifest themselves as a suppression
in the transverse momentum ($p_t$) distribution at a scale which is connected with the
rapidity of the measured particles. Saturation effects are most evident at large
(pseudo)-rapidity, i.e., at small angles relative to the beam direction. The dependence 
of the nuclear modification factor on the number of participants provides a constraint
 on the mechanism underlying the high-$p_t$ suppression. Here we present results for the 
nuclear modification factor $R_{cp}$ at forward rapidity as a function of system mass
comparing results from CuCu, AuAu and dAu collsions at $\sqrt{s_{NN}} = 200$ GeV. 

\end{abstract}

\maketitle

The main goal of the heavy ion program at the relativistic heavy ion collider(RHIC)
 is the creation and detection of the quark gluon plasma (QGP), a deconfined
 state of matter consisting of quarks and gluons(partons). High-$p_t$ partons going through
a QGP are expected to suffer significant energy loss resulting in fewer high-$p_t$ particles
in the exit channel than would be present without the QGP \cite{BSZ}. 
An experimental observation of such a supression in high energy heavy ion collisions
 would provide evidence that a QGP has been created. In contrast, the ratio of dAu to 
pp-collisions is expected to show an enhancement over a range in momentum, a phenomennon known 
as the Cronin effect \cite{GVWZ}. The argument is that energy loss in the "cold matter" 
of dAu collisions is quite small and the Cronin effect is a result of $p_t$ broadening due 
to multiple scattering. On the other hand, if a suppression is observed in dAu colisions at
forward rapidities instead of the Cronin effect, it may be an indication that initial state
 effects are important. A marked high-$p_t$ suppression with increasing pseudorapidity may
 thus be related to the initial conditions of the colliding deutron and gold nuclei, in 
particular to the possible existence of the color glass condensate(CGC)\cite{GLR,KLM}.

The suppression of high-$p_t$ hadron yields has commonly been investigated
 using the nuclear modification factor $R_{AA}$ which is the ratio of the measured hadron
 spectra to a scaled reference spectra from pp collisions. The value of $R_{AA}$ should be unity
 if particle production scales with the average number of binary nucleon-nucleon collisions $N_{coll}$.
 In the absence of a good pp reference, it is also possible to measure hadron suppression in terms of 
$R_{cp}$,the ratio of central to peripheral data scaled by $<N_{coll}>$. If there are no
 medium effects, this ratio is also expected to be equal to unity.

\begin{equation}
R_{cp}= \frac{Yield^{central}/<N_{coll}^{central}>}{Yield^{peripheral}/<N_{coll}^{peripheral}>} \nonumber
\end{equation}

We present here preliminary results on $R_{cp}$ in CuCu Collisions at 200 GeV measured at
 the BRAHMS experiment. The BRAHMS detector system \cite{ADAM} consists of global detectors
 for event characterization, a Mid-Rapidity Spectrometer (MRS) and a Forward Spectrometer (FS)
 covering forward rapidities. A multiplicity array(MA) consisting of scintillator tiles and
 silicon strip detectors mounted coaxially around the beam axis was used to characterize the
 centrality of collisions. For the present studies, the forward spectrometer was positioned
 at $4^o$ with respect to the beam direction, corresponding to an $\eta$ range between 
2.9 and 3.4. Events within $\pm20$ cm of the nominal interaction vertex were considered.

The motivation for the CuCu system was that it serves as a bridge between dAu and AuAu 
collisions in terms of the number of participants ($N_{part}$) and number of binary collisions.
 In particular, looking at the data in terms of $N_{part}$ offers the possibility of studying
 the system size dependence of the nuclear modification factor, and may help disentangle the
 initial state versus final state scenarios put forward to explain the high $p_t$ results at RHIC. 

 Experiments at RHIC have revealed that the matter produced in high energy heavy ion 
collisions is a strongly interacting dense medium , or sQGP. Results from Au+Au 
collisions at $\sqrt{s_{NN}}$ = 130 and 200 GeV have shown that hadron production at
 these energies is strongly suppressed relative to expectations based on an independent
 superposition of nucleon-nucleon collisions at $p_t$ of 2-10 GeV.

The BRAHMS experiment has already measured the transverse momentum spectra of hadrons 
both at mid-rapidity and at forward rapidities in dAu and Au+Au collisions \cite{BRAHMS1}. 
The absence of high $p_t$ suppression at mid-rapidity in dAu  collisions is 
consistent with the hypothesis of parton energy loss or/and parton recombination in the 
dense medium formed in Au+Au collisions indicating that the suppression may be due to 
final-state effects. However, forward rapidity results showed a suppression in
 dAu collisions suggesting that not only final-state effects but also initial-state effects
 such as CGC may also play a role.

\begin{figure}
\includegraphics[width=0.9\textwidth,height=0.5\textwidth]{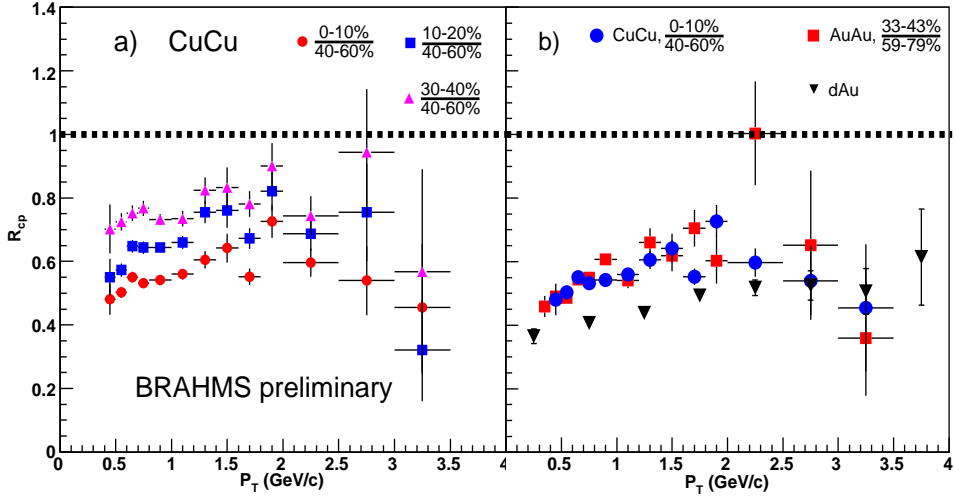}
\caption{$R_{cp}$ at $\eta = 3.2$ in CuCu Collisions as a function of $p_t$,(a) Centrality dependence, (b) comparison to results from dAu and AuAu Collisions. For CuCu, $0-10\%$ corresponds to $N_{part}\approx 96.8$ and $40-60\%$ to $N_{part}\approx 18.7$. For AuAu $33-43\%$ corresponds to $N_{part}\approx 96.3$ and $59-79\%$ to $N_{part}\approx 17.8$. \label{Rcp_CuCu_Comp}}
\end{figure}

Figure \ref{Rcp_CuCu_Comp}(a) shows the ratio $R_{cp}$ of yields from CuCu collisions, at $\eta = 3.2$ ,
 of a given centrality class to yield from the most peripheral collisions(40-60\%) scaled by the
 number of binary collisions in each sample. The data for the different centrality classes are 
obtained from the same collider run. As a result, the ratios are largely free of systematic
 errors associated with run-by-run collider and detector performance. The dominant systematic
 error in the $R_{cp}$ ratios come from the determination of $N_{coll}$ in the centrality bins. One can
 see that the $R_{cp}$ increases for more central collisions.

Figure \ref{Rcp_CuCu_Comp}(b) shows $R_{cp}$ in CuCu in comparison with those measured in dAu and AuAu 
collisions at the same pseudo-rapidity. It is evident that the suppression in CuCu follows a 
similar trend as in dAu and AuAu. An important question is whether $R_{cp}$ values in 
CuCu and AuAu are similar if one considers the case where the mean number of participants at
 the same collision energy are equal. As can be seen from the plots, $R_{cp}$ is very similar in both 
CuCu and AuAu collisions for a similar number of participants. 

In summary, we have presented preliminary results on $R_{cp}$ in CuCu collisions at 
$\sqrt{s_{NN}}$ = 200 GeV and $\eta = 3.2$. Our results show that one obtains similar
 results for $R_{cp}$ considering the same number of participants in both CuCu 
and AuAu collisions. The implication here may be that what matters is the geometry of the
 interaction region, i.e., the volume through which produced hadrons have to travel. 
To complete the system mass systematics, we are currently extending these studies to the
 midrapidity region and to the lower energy of $\sqrt{s_{NN}}$ = 62 GeV.

\begin{theacknowledgments}
This work was supported by the office of Nuclear Physics of the U.S. Department of energy,
the Danish Natural Science Research Council, the Research Council of Norway, the Polish 
State Committe for Scientific Research (KBN), and the Romanian Ministry of Research.
\end{theacknowledgments}

\end{document}